\documentclass{nature}

\usepackage{graphicx}
\usepackage{dcolumn}
\usepackage{bm}
\usepackage{color}
\usepackage{float}
\usepackage{sistyle}
\usepackage{amsmath}
\usepackage{mathtools}
\usepackage{subfigure}  
\usepackage{verbatim}   
\usepackage{lineno}
\usepackage{ulem}

\makeatletter
\let\saved@includegraphics\includegraphics
\AtBeginDocument{\let\includegraphics\saved@includegraphics}
\renewenvironment*{figure}{\@float{figure}}{\end@float}
\makeatother

\title{Terahertz pulse-driven collective mode in the nematic superconducting state of Ba$_{1-x}$K$_x$Fe$_2$As$_2$} 

\begin{document}

\author{Romain Grasset$^{1,2,3}$, Kota Katsumi$^{2}$, Pierre Massat$^4$, Hai-Hu Wen$^5$, Xian-Hui Chen$^6$, Yann Gallais$^{1,2,4}$ \& Ryo Shimano$^{1,2}$}

\maketitle

\begin{affiliations}
\item Cryogenic Research Center, The University of Tokyo, Tokyo, 113-0032, Japan
\item Department of Physics, The University of Tokyo, Tokyo, 113-0033, Japan
\item Laboratoire des Solides Irradi\'es, Ecole Polytechnique, CNRS, CEA, Institut Polytechnique de Paris, F-91128 Palaiseau, France
\item Laboratoire Mat\'eriaux et Ph\'enom\`enes Quantiques, UMR 7162 CNRS, Universit\'e de Paris, B$\hat{a}$t. Condorcet 75205 Paris Cedex 13, France
\item National Laboratory of Solid State Microstructures and Department of Physics, Nanjing University, Nanjing 210093, China
\item Hefei National Laboratory for Physical Sciences at Microscale and Department of Physics, and CAS Key Laboratory of Strongly-coupled Quantum Matter Physics, University of Science and Technology of China, Hefei, Anhui 230026, China 
\end{affiliations}

\section*{Abstract}
\begin{abstract}
We investigate the collective mode response of the iron-based superconductor Ba$_{1-x}$K$_x$Fe$_2$As$_2$ using intense terahertz (THz) light. In the superconducting state a THz Kerr signal is observed and assigned to non-linear THz coupling to superconducting degrees of freedom. The polarization dependence of the THz Kerr signal is remarkably sensitive to the coexistence of a nematic order. In the absence of nematic order the $C_4$ symmetric polarization dependence of the THz Kerr signal is consistent with a coupling to the Higgs amplitude mode of the superconducting condensate. In the coexisting nematic and superconducting state the signal becomes purely nematic with a vanishing $C_4$ symmetric component, signaling the emergence of a new superconducting collective mode activated by nematicity. 
\end{abstract}

\maketitle
\section*{Introduction}
Superconductivity with coexisting electronic orders can be found in various strongly correlated systems. Among these orders electron nematics, where the electron fluid breaks the discrete rotational symmetry of the underlying lattice, have recently emerged as an ubiquitous phase in many superconductors ranging from cuprates \cite{fradkin_colloquium_2015}, to iron-based superconductors \cite{fernandes_what_2014} where superconductivity emerges within a nematic phase, and more recently doped Bi$_2$Se$_3$ \cite{yonezawa_thermodynamic_2017} and twisted bi-layer graphene \cite{kerelsky_maximized_2019,cao_nematicity_2021} where the superconductivity itself may have a nematic component. In iron-based superconductors (Fe SC), superconductivity is found to coexist with both stripe-like magnetic spin-density-wave (SDW) and nematic orders. BaFe$_2$As$_2$, a member of this family, undergoes a nematic-structural transition from a $C_4$ to a $C_2$ symmetric phase, followed by a SDW transition \cite{canfield_feas-based_2010,fernandes_what_2014}. The $C_4$ rotational symmetry breaking is triggered by electronic degrees of freedom and has been dubbed nematic for this reason \cite{chu_divergent_2012,gallais_observation_2013,fernandes_what_2014}. With increasing doping by substitution (e.g. Ba with K or Fe with Co \cite{canfield_feas-based_2010,bohmer_superconductivity-induced_2015}), the $C_2$ symmetric nematic-SDW phase, hereafter called the $C_2$ phase, is weakened and a superconducting (SC) dome forms around a possible quantum critical point. The coexistence with the $C_2$ phase can profoundly impact the nature of SC order, by coupling different nearly degenerate pairing channels like $s$ and $d$ wave \cite{graser_near-degeneracy_2009,bohm_microscopic_2018}, or inducing an orbitally-selective SC state \cite{fernandes_nematicity_2013,sprau_discovery_2017}.   
\par

One way to gain insight into the coupling between nematic and SC degrees of freedom is to study the collective modes of the SC state upon entering the $C_2$ SC phase. Theoretically, intertwined electronic orders where superconductivity coexists with other electronic orders can lead to a rich spectrum of SC collective modes \cite{littlewood_amplitude_1982,moor_dynamics_2014,fu_quantum_2014,raines_hybridization_2015,dzero_amplitude_2015,gallais_nematic_2016,soto-garrido_higgs_2017,fu_quantum_2014,sentef_theory_2017,morice_collective_2018,muller_collective_2019,muller_interplay_2020}. In a single band conventional superconductor the collective mode excitation spectrum consists of two modes: the Nambu-Goldstone phase mode which is shifted to the plasma frequency through the Anderson-Higgs mechanism, and the Higgs amplitude mode located at twice the SC gap energy. The Higgs mode does not couple linearly to light \cite{littlewood_amplitude_1982,pekker_amplitude_2015,shimano_higgs_2020}. Except in very special cases like charge-density-wave superconductors \cite{sooryakumar_raman_1980,measson_amplitude_2014,grasset_higgs_2018,grasset_pressure_2018}, its observation has remained elusive until very recently. 
In this context, strong terahertz (THz) pulses have emerged as a tool of choice because they can access hidden SC collective modes via non-linear optical processes \cite{tsuji_theory_2015,cea_nonlinear_2016,murotani_nonlinear_2019,udina_theory_2019,tsuji_higgs-mode_2020,seibold_third_2020,schwarz_theory_2020,tsuji_higgs-mode_2020}. This has led to the observation of the SC Higgs mode in several SC materials like NbN and Nb$_3$Sn, but also in cuprates and Fe SC \cite{matsunaga_higgs_2013,matsunaga_light-induced_2014,matsunaga_polarization-resolved_2017,nakamura_infrared_2018,katsumi_higgs_2018,chu_new_2019,yang_lightwave-driven_2019,vaswani_light_2020}. In the case of Fe SC however, little is known experimentally about the impact of nematicity on SC collective modes like the Higgs.

\par
Here we investigate the THz non-linear response of the Fe SC Ba$_{1-x}$K$_x$Fe$_2$As$_2$ where superconductivity coexists with a nematic order using a THz pump near-infrared (NIR) probe scheme. In the SC state, we observe an instantaneous response which follows the square of the THz electric field which is assigned to the non-linear THz Kerr effect. In the absence of a coexisting nematic order the THz Kerr signal displays a $C_4$ symmetric polarization dependence consistent with a non-linear coupling to the SC Higgs mode. In the presence of a coexisting nematic order, the THz Kerr signal displays a drastic change in its polarization dependence: from fully-symmetric in the $C_4$ symmetric SC phase to fully nematic in the $C_2$ symmetric phase. We show theoretically that the onset of the THz Kerr Higgs response in the nematic channel can be qualitatively explained by taking into account the anisotropy of the electronic structure in the $C_2$ nematic phase. However, the complete disappearance of the $C_4$ symmetric signal in the $C_2$ SC phase cannot be captured within this simple picture, indicating a non-trivial interplay between the nematic and superconducting order parameters and the emergence of a new collective mode, distinct from the Higgs mode. We tentatively assigned this mode to the Bardasis-Schrieffer mode connecting s-wave and d-wave superconducting ground states which become mixed in the $C_2$ symmetric SC phase.

\section*{Results}
\begin{figure}[ht!]
		\begin{center}
		\includegraphics[width=13cm]{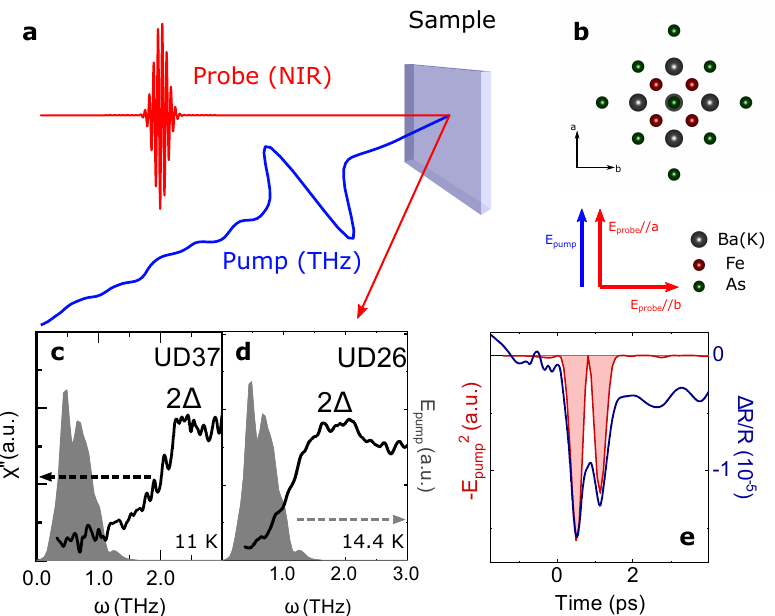}
		\caption{\textbf{Non linear THz response in Ba$_{1-x}$K$_x$Fe$_2$As$_2$.} (a) Sketch of the THz pump near-infrared (NIR) probe measurements. (b) Crystal structure of BaFe$_2$As$_2$ and the two polarization configurations used to determine the $C_4$ symmetric and nematic components of the transient reflectivity. (c,d) Raman spectra of Ba$_{1-x}$K$_x$Fe$_2$As$_2$ in the B$_{1g}$ symmetry for UD37 (c) and UD26 (d) below $T_c$. $2\Delta$ indicates superconducting gap from the hole  pockets. The grey curve represents the energy spectrum of the THz pump. (e) For UD37, $\Delta$R/R (blue line) along one of the Fe-Fe axis at 20~K ($T<T_c$) as a function of delay time between the pump and probe pulses. The E$_{pump}^2$ (red line) component corresponds to the THz Kerr effect.}
		\label{f:1}
		\end{center}
	\end{figure}

\textbf{Non-linear THz Kerr effect.} 
We studied two single crystals of Ba$_{1-x}$K$_x$Fe$_2$As$_2$ with $T_c=26$K (UD26) and $T_c=37$~K (UD37). The UD26 crystal is slightly underdoped and exhibits a simultaneous nematic/SDW transition at $T_{N} \sim T_S \sim 90$~K. The UD37 crystal only exhibits a superconducting transition and is close to optimal doping. The terahertz-pump optical reflectivity probe (TPOP) measurement scheme is depicted in Figure \ref{f:1}(a). Measurements were carried out with a fixed THz pump polarization along the Fe-Fe direction but two different probe polarizations either parallel or perpendicular to the pump polarization (Figure \ref{f:1}(b)). Fe-Fe directions are identified by a 45° tilt with respect to the edges of the crystals which are square-shaped. 
\par
In Figure \ref{f:1}(c,d) we compare the THz pump spectrum with the SC state Raman spectra of the two samples \cite{SM}. With an energy centered around $\omega_p$=0.6~THz=20~cm$^{-1}$, the THz pump spectrum is located below the lowest superconducting gap $2\Delta_h$ observed by Raman scattering. Based on previous Raman and angle-resolved photoemission spectroscopy (ARPES) measurements this gap is assigned to the $\Gamma$ centered hole pockets. The TPOP signal $\frac{\Delta R}{R}$ of UD37 below $T_c$ is shown in Fig. \ref{f:1}(e). It consists of essentially two components, an instantaneous component that follows the square of the THz $E$-field (red line in Fig. \ref{f:1}(e)) and a broader decaying component which last several picoseconds after the pump pulse. In the following we will mostly focus on the instantaneous component, the THz Kerr effect, where the strong THz $E$ field modulates the optical reflectivity in the NIR regime \cite{hoffmann_terahertz_2009}. We note that in our measurements we only detect an instantaneous component that is proportional to the square of the THz $E$ field, consistent with the centrosymmetric crystal structure of Ba$_{1-x}$K$_x$Fe$_2$As$_2$. No forbidden odd contribution is observed, as recently reported in the SC state of NbN\cite{Yang2019} and attributed to THz field symmetry breaking.
\par
The THz Kerr signal is described by a third-order nonlinear susceptibility $\chi^{(3)}(\omega;\omega,+\Omega,-\Omega)$
\cite{boyd_nonlinear_2008,paul2021}, where $\omega$ and $\Omega$ are the frequencies of the NIR pulse and THz-pump pulse, respectively. The THz pulse-induced reflectivity change $\Delta R/R$ can be expressed in terms of $\chi^{(3)}$ \cite{SM} as
 \begin{equation}
 \frac{\Delta R}{R}(E^{probe}_i,E^{probe}_j)\sim \frac{1}{R}\frac{\delta R}{\delta \epsilon_1}\epsilon_0 [Re\chi^{(3)}_{ijkl}]E^{pump}_kE^{pump}_l
 \label{e:1}
 \end{equation}
where $E_i$ denotes the $i$th component of the THz-pump or probe $E$ field and $\epsilon_1$ is the real part of the dielectric constant at 1.5 eV.
The instantaneous Kerr signal of interest here implies $\Omega$=0 in $\chi^{(3)}$. It is therefore independent of the pump frequency $\Omega$ and non-resonant \cite{paul2021,katsumi_higgs_2018}. This is in contrast with the Third-Harmonic Generation (THG) signal which is resonant when the pump frequency $\Omega$ equals the superconducting gap $\Delta$ \cite{tsuji_theory_2015}.
\par
In general, the onset of a THz Kerr signal below $T_c$ can be assigned to two different processes: coupling to charge density fluctuations (CDF) like the one observed in Raman experiments, or to the SC Higgs mode. As previously shown in the case of NbN and cuprates important clues about the origin of the THz Kerr signal, and other third-order non-linear effects like THG, can be obtained by investigating its polarization dependence \cite{cea_nonlinear_2016,matsunaga_polarization-resolved_2017,katsumi_higgs_2018,chu_new_2019,tsuji_higgs-mode_2020,seibold_third_2020}.

Assuming $C_4$ tetragonal symmetry for the normal state of Ba$_{1-x}$K$_x$Fe$_2$As$_2$, we can analyze the polarization dependence of $\chi^{(3)}(\theta_{pump},\theta_{probe})$ in terms of the irreducible representations of $D_{4h}$ point group as
\begin{equation}
\begin{multlined}
\chi^{(3)}(\theta_{pump},\theta_{probe})=\frac{1}{2}(\chi^{(3)}_{A_{1g}}+\chi^{(3)}_{B_{1g}}\textrm{cos}2\theta_{pump}\textrm{cos}2\theta_{probe} \\
 +\chi^{(3)}_{B_{2g}}\textrm{sin}2\theta_{pump}\textrm{sin}2\theta_{probe})
\end{multlined}
\end{equation}
where we have defined the symmetry-resolved non-linear response functions: $\chi^{(3)}_{A_{1g}}=\chi^{(3)}_{aaaa}+\chi^{(3)}_{bbaa}$, $\chi^{(3)}_{B_{1g}}=\chi^{(3)}_{aaaa}-\chi^{(3)}_{bbaa}$ and $\chi^{(3)}_{B_{2g}}=\chi^{(3)}_{abab}+\chi^{(3)}_{abba}$.
 and $\theta_{probe/pump}$ are the angle between the probe/pump polarization vectors and the $a$ axis of the 1 Fe unit cell. The $A_{1g}$ is the fully symmetric representation and the $B_{1g}/B_{2g}$ representation transform as $x^2-y^2$ and $xy$ respectively. The $B_{1g}$ representation has the same symmetry as the $C_2$ symmetric nematic order parameter found in Fe SC.
For $\theta_{pump}$=0, the $A_{1g}$ and $B_{1g}$ responses can be accessed using two distinct probe polarization orientations. Indeed making use of equation \ref{e:1} we can write: 
\begin{equation}
\begin{multlined}
\frac{\Delta R}{R}^{C_4}= \frac{\Delta R_a}{R_a}+\frac{\Delta R_b}{R_b} \propto Re\chi^{(3)}_{A_{1g}} \\
\frac{\Delta R}{R}^{nem}= \frac{\Delta R_a}{R_a}-\frac{\Delta R_b}{R_b} \propto Re\chi^{(3)}_{B_{1g}}
\end{multlined}
\end{equation}
where $R_i$ ($i=a,b$) denotes the reflectivity for a probe polarization along the Fe-Fe axes (a,b) of Figure \ref{f:1}.b and for a fixed pump polarization along the a axis. Here we have taken $R_a=R_b$ ($C_4$ tetragonal symmetry). Since the notations $B_{1g}$ and $A_{1g}$ are no longer valid in the $C_2$ symmetric orthorhombic phase, we will adopt the notation "$C_4$" for $C_4$ symmetric and "nem" for nematic (or $C_2$ symmetric) when discussing the results below.
\par
\begin{figure}[tb!]
		\begin{centering}
		\includegraphics[width=17cm]{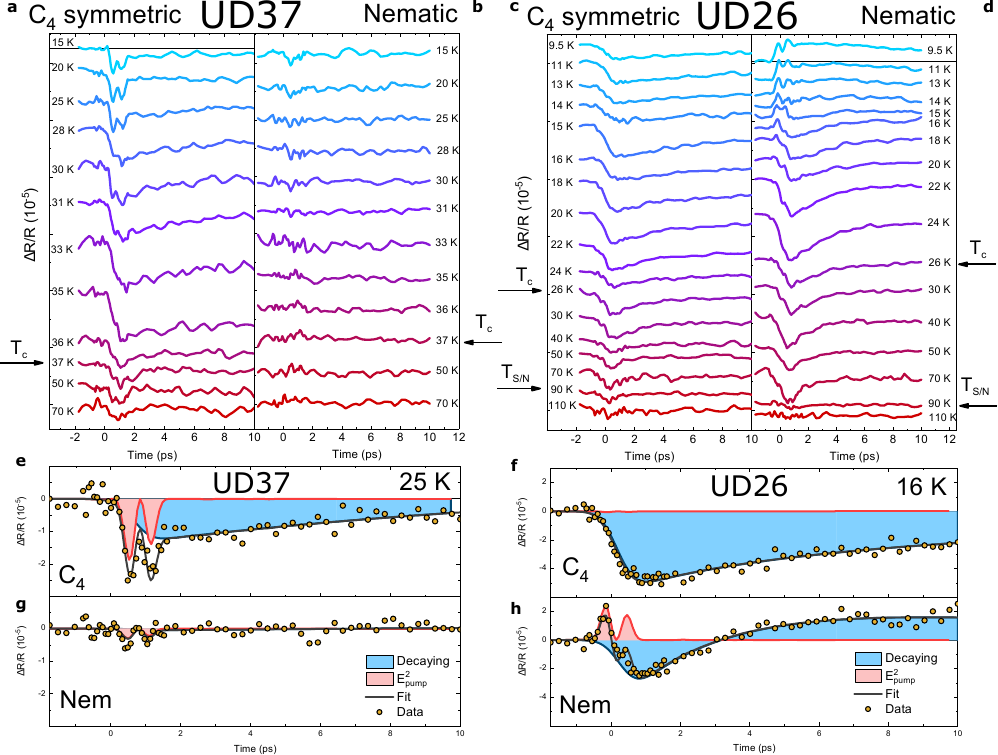}
		\caption{\textbf{Temperature and symmetry dependence of the THz response.} (a-d)$\Delta$R/R against the delay time in the $C_4$ symmetric ($C_4$) and nematic (Nem) channels at various temperatures for UD37 (a,b) and UD26 (c,d). (e-h) Fitted curves for the different components of $\frac{\Delta R}{R}$ at $T<T_c$ for UD37 (e,g) and UD26 (f,h).}
		\label{f:2}
		\end{centering}
	\end{figure}
\begin{figure}[tb!]
		\begin{center}
		\includegraphics[width=11cm]{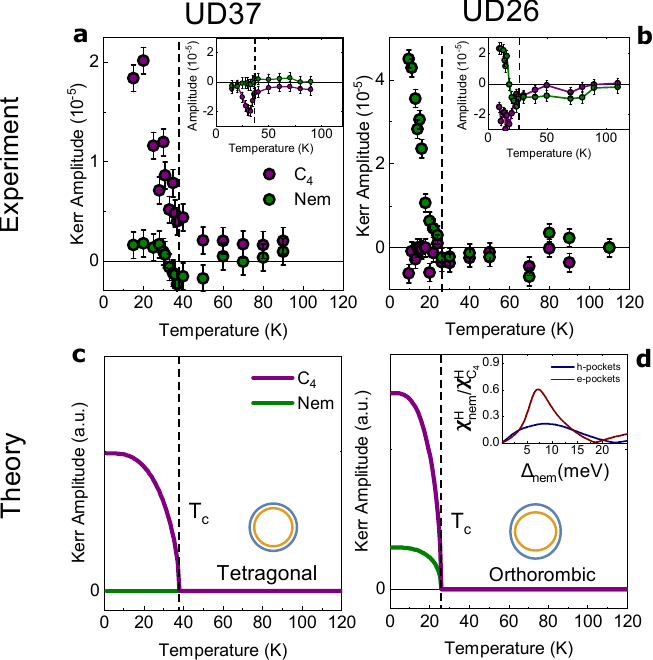}
		\caption{\textbf{Instantaneous Kerr response} (a,b) Temperature dependencies of the amplitude of the fitted instantaneous Kerr signals in the $C_4$ symmetric ($C_4$) and nematic (nem) channels for UD37 (a) and UD26 (b). Insets: Temperature dependencies of the amplitude of the fitted SC decaying signals. (c,d) Calculated temperature dependence of Higgs contribution of the hole pockets to the instantaneous Kerr signals in the $C_4$ symmetric ($C_4$) and nematic (nem) channels for a tetragonal (c) and orthorombic (d) symmetry of the electronic dispersion relation. The shape of the Fermi surfaces of our model is represented in blue and yellow. Inset: Ratio of the nematic and $C_4$ symmetric components of the Higgs response as a function of the nematic order parameter for the hole (blue) and electron pockets (red).}
		\label{f:3}
		\end{center}
	\end{figure}

\textbf{Response in the $C_4$ symmetric superconducting state.} 
We start by discussing the $C_4$ symmetric and nematic components of the TPOP signal of the UD37 crystal for which only superconductivity is present. Figure \ref{f:2}(a,b) show the transient reflectivity obtained for both $C_4$ symmetric and nematic components in the UD37 crystal and at various temperatures ranging from 15~K to 70~K. The decaying part of the $\Delta R/R$ signal shows a strong increase in the $C_4$ symmetric channel across $T_c$ indicating the superconducting transition (Inset of Figure \ref{f:3}(a)).  The instantaneous Kerr component is only observed below $T_c$, confirming it is linked to the onset of superconductivity. On the other hand, in the nematic channel no significant changes appear in the transient reflectivity at all temperatures. 
Using the fitting procedure displayed in Figure \ref{f:2}(e,g), we can obtain the temperature dependencies of the instantaneous and decaying components of $\frac{\Delta R}{R}^{C_4}$ and $\frac{\Delta R}{R}^{nem}$ (Figure \ref{f:3}(a)). The decaying component displays a sharp extremum around $T_c$ and is assigned to the dynamical relaxation of quasi-particles (QP) in the SC state \cite{SM}. The instantaneous component, attributed to the THz Kerr effect, shows a strong enhancement below $T_c$. The absence of the instantaneous Kerr component in the nematic channel argues in favor of a contribution arising from the Higgs excitation. Indeed, as shown in the case of Bi$_2$Sr$_2$CaCu$_2$O$_{8+x}$ cuprates charge density fluctuations are expected to contribute to all symmetry channels whereas the Higgs contribution is only active in the fully-symmetric, i.e. $C_4$ symmetric for a tetragonal crystal, channel \cite{matsunaga_polarization-resolved_2017,cea_polarization_2018,katsumi_higgs_2018,seibold_third_2020,tsuji_higgs-mode_2020}. Interestingly, Raman scattering spectra on the same crystal in the SC state are dominated by the $B_{1g}$ channel \cite{bohm_microscopic_2018,SM}, indicating they mostly probe CDF contributions in stark contrast with the THz Kerr signal. We note that the respective weight between Higgs and CDF contributions to third-order non-linear susceptibilities has been a subject of a debate since BCS calculations indicate dominant CDF contributions for a clean superconductor \cite{cea_nonlinear_2016}. However, there is now an emerging consensus that disorder significantly boosts the Higgs contribution, thus giving a rationale to the dominance of the Higgs contribution observed in THz Kerr and THG experiments in all superconductors studied so far \cite{silaev_nonlinear_2019,jujo_quasiclassical_2018,murotani_nonlinear_2019,seibold_third_2020,tsuji_higgs-mode_2020}.
\par

\textbf{Response in the $C_2$ symmetric superconducting state.} 
Having discussed the simple case of the $C_4$ symmetric superconductor case, let us now turn to the sample with a lower doping level, UD26 which display a $C_2$ symmetric SC phase with both nematic and SC orders. Figure \ref{f:2}(c,d) shows the transient reflectivity obtained for both channels and at various temperatures ranging from 9.5~K to 110~K. Above $T_c$, in contrast to UD37, both $C_4$ symmetric and nematic components show a change in the transient reflectivity below $T_{S/N}\sim 90$~K indicating the transition to the $C_2$ symmetric nematic phase. The onset of a decaying signal in the nematic channel is consistent with optical pump optical probe measurements on BaFe$_2$(As$_{1-x}$P$_x$)$_2$ which reported a similar strongly anisotropic signal below $T_{S/N}$ \cite{thewalt_imaging_2018}. In principle, a mixture of $C_2$ domains of different orientation would average out the nematic component of our signal. The fact that we observe a significant non-zero $\frac{\Delta R}{R}^{nem}$ shows that one domain orientation prevails under our 250~$\mu$m laser spot. We attribute this relatively large domain size to residual strains on the sample due to sample mounting which act as symmetric breaking field and align the nematic domains.
\par
Below $T_c$, an instantaneous Kerr component of $\Delta R/R$ that follows the squared THz-pump E-field is also identified, with however a striking difference compared to UD37. Indeed, while it is essentially absent in $\frac{\Delta R}{R}^{C_4}$, the instantaneous Kerr signal shows a strong enhancement below $T_c$ in the $\frac{\Delta R}{R}^{nem}$ channel. 
Using the fitting procedure displayed in Figure \ref{f:2}(f,h), we obtained the amplitude of the instantaneous Kerr and decaying components (See Figure \ref{f:3}(b) and \cite{SM}). Interestingly, the channel dependence of the instantaneous Kerr and decaying signal are distinct: while the instantaneous Kerr signal is fully nematic with no $C_4$ symmetric component within our experimental accuracy, the decaying signal is present in both channels with similar amplitudes at all temperatures \cite{SM}.
\par
\textbf{Origin of the nematic response} 
The complete switch from $C_4$ symmetric to nematic channel of the instantaneous Kerr signal when going from the $C_4$ SC phase to the $C_2$ SC phase is the central finding of the present work. It indicates an unanticipated and profound impact of the $C_4$ symmetry breaking on the THz Kerr non-linear optical signal of the SC state. We now explore different scenarios to explain this phenomena. First since the structural transition from tetragonal to orthorhombic involves a mixing of the A$_{1g}$ and B$_{1g}$ symmetry into the A$_g$ symmetry, we naturally expect some mixing of the $\frac{\Delta R}{R}$ symmetry components due to the anisotropy of the optical constants. Based on optical measurements on detwinned BaFe$_2$As$_2$ samples \cite{nakajima_unprecedented_2011,SM}, we determined quantitatively how the two symmetries are mixed from the calculation of the $\frac{1}{R}\frac{\delta R}{\delta \epsilon_1}$ 
pre-factor in Equation \ref{e:1}. We found at most a 25$\%$ anisotropy with respect to the a and b axes. As expected this anisotropy causes a non-zero nematic component. However, it leads to a $\frac{\Delta R}{R}^{nem}$ signal of at most 10$\%$ of the $\frac{\Delta R}{R}^{C_4}$ signal, and therefore plays a marginal role in the $C_4$ symmetric to nematic transition observed in the THz Kerr signal. 
\par
Having ruled out a simple effect of anisotropic linear optical constant, we are left with the properties of the non-linear response $\chi^{(3)}$ itself. In the $C_2$ phase, the anisotropy of the electronic dispersion relation will also induce a non-zero component of the Higgs mode response in the nematic channel \cite{schwarz_theory_2020}. We evaluated the activation of the Higgs response in the nematic channel by calculating the third-order non-linear Higgs response (See Fig. \ref{f:3}(c,d) for the contribution of the hole pockets) using a three pocket model (2 hole-like and 1 electron-like) and an s-wave superconducting gap \cite{SM}. As expected, in the $C_4$ phase the Higgs response appears below $T_c$ only in the $C_4$ symmetric channel in agreement with our observations in the UD37 sample. In the $C_2$ phase however, the distorted Fermi pockets due to finite nematic order parameter activate the Higgs mode in the nematic channel as observed experimentally. The activation grows with the nematic splitting energy $\Delta_{nem}$, but quickly saturates and decreases (See inset of Fig. \ref{f:3}(d)). We found that for any realistic nematic splitting energy and band parameters, the nematic response of either hole or electron pockets is at most 60$\%$ of the $C_4$ symmetric response, thus failing to explain the experimental observation. We note that a dominant contribution from CDF to the THz would be inconsistent with both the fully $C_4$ symmetric Kerr signal observed in UD37 and the fully nematic Kerr signal observed in UD26 (see \cite{SM} for an evaluation of the CDF contribution).

\section*{Discussion}
From the above discussion, it appears that the strong dominance of the THz Kerr signal in the nematic channel of UD26 cannot be explained simply by the effect of the anisotropy of the optical constant or the electronic structure on the Higgs signal. We are thus left with more speculative scenarios. 
First, we discuss the possibility of the an exotic SC order parameter in the $C_2$ phase. We note that an SC order parameter with lower symmetry like $d$-wave will not by itself activate a Higgs Kerr signal in non-fully symmetric channels as demonstrated in the case of cuprates \cite{katsumi_higgs_2018,schwarz_theory_2020}, so that our observations cannot be easily linked to a change in SC gap symmetry at least for a single band superconductor. However, in multi-orbital systems like Fe superconductors it is possible that the internal structure of Cooper pairs in orbital space profoundly affects the anisotropy of the Kerr Higgs signal. An intriguing possibility is the recent proposal of an orbital-selective SC state in the $C_2$ phase of FeSe \cite{sprau_discovery_2017}. Whether such state would by itself yield a Higgs signal in the nematic channel only is unclear and deserves further theoretical investigations.
\par
Another possibility is that the THz Kerr signal arises from an SC collective mode which couples to the nematic order parameter. In Ba$_{1-x}$K$_x$Fe$_2$As$_2$ $s$ and $d$ wave pairing channels are close competitors, potentially giving rise to a Bardasis-Schrieffer (BS) mode in the nematic $d$ wave channel \cite{graser_near-degeneracy_2009, bardasis_excitons_1961, barlas_amplitude_2013}. Several spectral features of the Raman spectrum of Ba$_{1-x}$K$_x$Fe$_2$As$_2$ have indeed been interpreted as BS modes, consistent with theoretical evaluations of pairing instabilities in hole-doped BaFe$_2$As$_2$ \cite{graser_near-degeneracy_2009,bohm_microscopic_2018,maiti_collective_2015,maiti_probing_2016} (see also SM for a discussion of the Raman spectra in the SC state). Recently, Muller et al. have argued that in the $C_2$ phase the BS mode will couple to the amplitude mode of the nematic order parameter, giving raise to a single coupled nematic-BS mode below the Higgs mode energy due to the appearance of a strongly mixed $s$+$d$ SC state\cite{muller_interplay_2020}. The stronger decaying signal in the UD26 sample compared to the UD37 sample well-below $T_c$ supports the idea of an increased anisotropy of the SC gap in the $C_2$ SC phase in agreement with a significant $d$ wave admixture. Interestingly, for parameters close to the critical point where the nematic phase terminates Muller et al. found that the coupled nematic-BS mode may become dominant over the Higgs mode in the short times dynamics after a quench \cite{muller_interplay_2020}. Furthermore, we note that the BS mode has the nematic $B_{1g}$ symmetry and will naturally give rise to a signal in the nematic channel \cite{muller_signatures_2021}.  A computation of the third-order non-linear susceptibility taking into account both $s$ and $d$ pairing channels in the presence of a finite nematic order parameter is desirable to further assess this scenario.
\par
In conclusion, we have studied the impact of nematicity on the SC collective modes in Ba$_{1-x}$K$_x$Fe$_2$As$_2$ via THz pump optical probe measurements. In the absence of nematicity we observe an instantaneous behavior of the optical reflectivity which we assign to a THz Kerr coupling to the Higgs mode. In the coexisting nematic + SC phase we observe a drastic change in the polarization dependence of the THz Kerr signal from purely $C_4$ symmetric to purely nematic. The change cannot be accounted by the anisotropy of the electronic properties and indicates the emergence of a new SC collective mode which couples strongly to the nematic order parameter. The exact identification of this mode requires further investigation, but we suggest the Bardasis-Schrieffer mode connecting nearly-degenerate $s$ and $d$ wave pairing ground states as a likely candidate.
\section*{Methods}
\textbf{Samples.} The two single crystals of Ba$_{1-x}$K$_x$Fe$_2$As$_2$ with $T_c=26$K (UD26, $x\sim$ 0.23) and $T_c=37$~K (UD37, $x\sim$ 0.28) were characterized by SQUID magnetometry, wavelength dispersing spectroscopy and Raman scattering measurements. The samples are square-shaped with sides of $\sim 5$~mm. The crystal orientations of both samples were confirmed by polarization-resolved Raman spectroscopy measurements.\\
\textbf{Terahertz pump-optical reflectivity probe (TPOP).} Strong single cycle THz pump pulses (0.3 - 1 THz) with electric field reaching up to 350~kV/cm are generated using optical rectification of 1.5 eV NIR pulses in a LiNbO$_3$ crystal using the tilted pulse front technique \cite{hebling_generation_2008,watanabe_intense_2010}. For optical probe measurements 100 fs duration NIR pulses at 1.5~eV are used with a fluence of 10-100~nJ/cm$^2$ and a spot size of 250~$\mu m$ in diameter. The repetition rate of the NIR laser is 1kHz. 

\section*{Data availability}
The authors declare that [the/all other] data supporting the findings of this study are available within the paper [and its supplementary information files].
\section*{Code availability}
All the numerical codes that support the findings of this study are available from the
corresponding authors (R.G., Y. G. and R.S.) upon reasonable request.
\section*{Acknowledgements}
The authors would like to thank M. M\"uller, I. Eremin, R. Lobo, I. Paul and L. Benfatto for fruitful discussions. R. G. and Y. G. acknowledge the support from the Japan Society for the Promotion of Science.
\section*{Competing Interests:} the Authors declare no Competing Financial or Non-Financial Interests.
\section*{Author contributions}
The samples were grown by H-H.W. and X-H.C.; R.G. and K.K. performed THz pump optical probe experiments; P. M. performed Raman experiments; R.G. and Y.G. performed the calculations and analyzed data;  Y.G. and R. S. supervised the project; All authors discussed the results and wrote the manuscript.

\section*{Correspondence:} Correspondence should be addressed to R. Grasset~(email: romain.grasset@polytechnique.edu), Y. Gallais~(email:yann.gallais@u-paris.fr) and R. Shimano~(email: shimano@thz.phys.s.u-tokyo.ac.jp).

\section*{References}

\section*{Figure legends}

\end{document}